\documentclass[11pt]{article}
\usepackage[a4paper]{geometry}

\usepackage{array}
\usepackage{mathtools,multirow}
\usepackage[section]{placeins}
\usepackage{float,caption}
\usepackage{relsize}
\usepackage{makeidx,url,algorithmicx,algorithm,algpseudocode}
\usepackage{amssymb,amsmath,amsthm,color}
\usepackage{pdfsync,graphicx,subfigure,hyperref}
\usepackage{graphicx}
\usepackage{amsmath}
\usepackage[T1]{fontenc}
\usepackage[english]{babel}
\usepackage[utf8]{inputenc}
\usepackage{apacite}

\begin{document}
	
	\title{A Two-Dimensional Intrinsic Gaussian Markov Random Field for Blood Pressure Data}
	
	\author{
		Maria-Zafeiria Spyropoulou\thanks{Department of Mathematics, Statistics and Actuarial Science, University of Kent, Canterbury, CT2 7FS, UK.}, James Bentham\footnotemark[1]\\
		(Correspondence: {\tt mzs2@kent.ac.uk}.) 
	}
	\maketitle
	\begin{abstract}	
		\noindent  Many real-world phenomena are naturally bivariate. This includes blood pressure, which comprises systolic and diastolic levels. Here, we develop a Bayesian hierarchical model that estimates these values and their interactions simultaneously, using sparse data that vary substantially between countries and over time. A key element of the model is a two-dimensional second-order Intrinsic Gaussian Markov Random Field, which captures non-linear trends in the variables and their interactions. The model is fitted using Markov chain Monte Carlo methods, with a block Metropolis-Hastings algorithm providing efficient updates. Performance is demonstrated using simulated and real data.
	\end{abstract}
	\section{Introduction}
	Blood pressure is bivariate: systolic blood pressure (SBP) is a measurement of the force exerted by the heart when it beats, while diastolic blood pressure (DBP) is a measurement of resistance to blood flow in the blood vessels. Raised blood pressure (RBP) is defined as SBP of 140 mmHg or higher, or DBP of 90 mmHg or higher, and is estimated to affect more than one billion people worldwide (\cite{zhou2017worldwide}). RBP is a key risk factor for non-communicable diseases such as cardiovascular conditions, cancers and diabetes, which are responsible for approximately 70\% of global deaths each year (\cite{whofactsheet}).
	
	In the past decade, Bayesian hierarchical modelling has become an established technique in global health research. In particular, it has been used to make estimates at national level of mean values of risk factors including elevated blood pressure, excess body-mass index, diabetes, and sub-optimal lipid profiles (\cite{zhou2017worldwide}, \cite{abarca2017worldwide}, \cite{zhou2016worldwide}, \cite{ncd2020repositioning}). The models are sufficiently robust for the World Health Organization to report resulting estimates in publications such as the Global Diabetes Report (\cite{roglic2016global}). 
	However, while influential  
	(\cite{atlaschildob}, \cite{gnr2020}), these models only allow estimation of single variables at a time. Given that disease risk factors have complex interactions that vary over time and between countries, the existing models therefore fail to capture important information.
	
	Although recent and comprehensive estimates of trends in the mean values of SBP, DBP and RBP at national level are available (\cite{zhou2017worldwide}), there is limited understanding of how the interaction between SBP and DBP varies over time, between countries, and by age and sex. Here, we extend existing methodology (\cite{danaei2011national}, \cite{finucane2014bayesian}) to the two-dimensional case, specifying a Bayesian hierarchical model that allows SBP, DBP and their interaction to be estimated simultaneously. A key development is to extend the random walk (\cite{rue2005gaussian}), which modelled non-linear trends in individual variables, to two dimensions. To do so, we use a two-dimensional Intrinsic Gaussian Markov Random Field (IGMRF) (\cite{rue2005gaussian}, \cite{yue2010nonstationary}, \cite{thon2012bayesian}) as the prior distribution for the precision matrix of a two-dimensional random walk.
	
	The paper is organised as follows: Section 2 introduces our Bayesian hierarchical model, Section 3 describes the IGMRF in two dimensions, while Section 4 presents the implementation of the model. Section 5 provides the results of a simulation study, Section 6 describes an application to real-world data, while Section 7 summarises our findings and proposes future work.
	\section{Model and Assumptions}
	Data for DBP, SBP and their interaction are available from studies $i$ carried out in countries $j$ for age groups $h$. We model these data, $\mathbf{y}$, as shown in \eqref{one}, where the $D$, $S$ and $I$ subscripts refer to DBP, SBP and their interaction, respectively, and $j[i]$ refers to the country $j$ in which study $i$ was carried out.\\
	\begin{align}\label{one}
		\left[\begin{array}{c}
			y_{h,i,D}\\y_{h,i,S}\\y_{h,i,I}\\
		\end{array} \right] &=\left[\begin{array}{c}
			a_{j[i], D}\\a_{j[i], S}\\a_{j[i], I}\\ 
		\end{array} \right]
		+\left[\begin{array}{c}
			b_{j[i],D}t_i\\b_{j[i],S}t_i\\ b_{j[i],I}t_i
		\end{array} \right]
		+\left[\begin{array}{c}
			\mathbf{u}_{j[i],(D,S),t_{i}}
		\end{array} \right]
		+ \left[\begin{array}{c}
			X_{i,D}.\boldsymbol\beta\\ X_{i,S}.\boldsymbol\beta\\ X_{i,I}.\boldsymbol\beta 
		\end{array} \right]
		+\left[\begin{array}{c}
			\gamma_{i,D}(z_h)\\ \gamma_{i,S}(z_h) \\ \gamma_{i,I}(z_h)
		\end{array} \right]\notag\\
		&+\left[\begin{array}{c}
			e_{i,D}\\e_{i,S}\\ e_{i,I}
		\end{array} \right]
		+\left[\begin{array}{c}
			w_{h,i,D} \\w_{h,i,S}\\w_{h,i,I}
		\end{array} \right]
		+\left[\begin{array}{c}
			\epsilon_{h,i,D}\\\epsilon_{h,i,S}\\\epsilon_{h,i,I}
		\end{array} \right]
	\end{align}\\
	This model has the same components as the model of single variables (\cite{danaei2011national}, \cite{finucane2014bayesian}), but extends them to the two-dimensional case. It includes linear intercepts and slopes, $\mathbf{a}_{j[i]}$ and $\mathbf{b}_{j[i]}$, non-linear terms that vary over time, $\mathbf{u}_{j[i]}$, covariate effects $\mathbf{X}_i\boldsymbol\beta$, terms in age $\boldsymbol\gamma_i(z_h)$, study-specific random effects $\mathbf{e}_i$, age-varying study-specific random effects $\mathbf{w}_{h,i}$, and noise $\boldsymbol\epsilon_{h,i}$, assumed to be iid Gaussian. The model in two dimensions can therefore be expressed as: 
	\vspace{3mm}
	\begin{align} \label{norm}
		\textbf{y}_{h,i}&\sim \mathcal{N}(\textbf{a}_{j[i]}+\textbf{b}_{j[i]}t_i+\textbf{u}_{j[i],t_{i}}+\textbf{X}_i\boldsymbol\beta + \boldsymbol\gamma_i(z_h) + \textbf{e}_i, \textbf{SD}_{h,i}^2/n_{h,i} +\boldsymbol{\tau}_{i}^{2})
	\end{align} 
	\vspace{3mm}
	where
	\begin{align}\label{sd1}
		\textbf{SD}_{h,i}^2/n_{h,i}&=\left[\begin{array}{ccc}
			SD_{D,h,i}^2/n_{h,i},&SD_{S,h,i}^2/n_{h,i},&SD_{I,h,i}^2/n_{h,i}\end{array}\right]^T
	\end{align}
	and
	\begin{align}
		\boldsymbol\tau_i^2 &= \left[\begin{array}{ccc}
			\tau^2_{i,D},&\tau^2_{i,S},&\tau^2_{i,I}	
		\end{array}\right]^T
	\end{align}
	It can be seen in \eqref{one} and \eqref{norm} that there is a single non-linear sub-model $\boldsymbol{u}_{j[i],t_i}$, which provides simultaneous estimates of non-linear trends in SBP and DBP, as well as implicit estimates of trends in their interactions, using an IGMRF as described below.	
	\section{Intrinsic Gaussian Markov Random Fields}
	A random vector $\mathbf{x}	= (x_1,...,x_n)^T \in \mathcal{R}^n$ is called a Gaussian Markov Random Field (GRMF) linked to a labelled graph $ G=(V,E) $ with mean $\boldsymbol{\mu}$ and precision matrix $\textbf{Q}>0$ iff its density has the form 
	\begin{equation}\label{GMRF}
		\pi(\mathbf{x}) = (2\pi)^{-n/2}|\textbf{Q}|^{1/2}\exp\left(\frac{1}{2}(\textbf{x}-\boldsymbol{\mu})^T\textbf{Q}(\textbf{x}-\boldsymbol{\mu})\right)
	\end{equation} 
	and
	\begin{equation}\label{edge}
		Q_{ij}\neq 0 \Leftrightarrow \lbrace i,j\rbrace \in E, \quad\forall i\neq j
	\end{equation} 
	A GMRF can be any Gaussian distribution with a symmetric positive definite covariance matrix and vice versa. The properties of a GMRF are described by its precision matrix \textbf{Q}, since its structure indicates connections between nodes: as shown in \eqref{edge}, non-zero values in \textbf{Q} correspond to an edge in $G$. Where \textbf{Q} is dense, the graph is fully connected, but when the precision matrix is sparse, this implies conditional independence.
	
	IGMRFs are improper GMRFs, as they have a precision matrix that is not of full rank, and since the precision matrix is not of full rank, its inverse does not exist. This implies that IGMRFs do not have well-defined mean or covariance matrices. However, they do have the property that the mean of an IGMRF of order $T$ is defined up to the addition of a polynomial of order $T-1$ (\cite{rue2005gaussian}).  
	
	\subsection{One-dimensional IGMRFs}
	A full description of one-dimensional IGMRFs is provided elsewhere (\cite{rue2005gaussian}). For a first-order random walk, the rank of the precision matrix is $T-1$, where $T$ expresses the number of rows or columns, and the behaviour in the first and last rows is different from the rest of the matrix, as shown in \eqref{IGMRF1}. These are the boundary constraints, and in this case, imposing the constraint $\sum_i x_i=0$ on the mean of the intercepts gives a proper joint density (\cite{sorbye2014scaling}). For a second-order random walk, the rank of the precision matrix is $T-2$, and the behaviour in the first two and last two rows is different from the rest of the matrix, as shown in \eqref{IGMRF2}. There are now two constraints, $\sum_i x_i=0$ and $\sum_i ix_i=0$, corresponding to the means of the intercepts and slopes. 
	\begin{align}\label{IGMRF1}
		\mathbf{Q}_{T}= \lambda
		\left( \begin{array}{ccccc}
			1&-1&&&\\
			-1&2&-1&&\\ 
			&-1&2&-1&\\
			&&-1&2&-1\\
			&&&-1&1\\
		\end{array} \right)
	\end{align}
	
	\begin{align}\label{IGMRF2}
		\mathbf{Q}_{T}= \lambda
		\left( \begin{array}{ccccc}
			1&-2&1&&\\
			-2&5&-4&1&\\ 
			1&-4&6&-4&1\\ 
			&1&-4&5&-2\\
			&&1&-2&1\\ 
		\end{array} \right)
	\end{align}	
	\subsection{Two-dimensional IGMRFs}
	To estimate non-linear trends in DBP, SBP and their interactions simultaneously, we use an IGMRF with precision matrix $\textbf{Q}$ as a prior distribution at each level of the hierarchy, with $\textbf{u}_c$, $\textbf{u}_r$, $\textbf{u}_s$ and $\textbf{u}_g$ values that represent non-linearity at country, region, super-region and global level, respectively. We have
	\vspace{2mm}
	\begin{align}\label{four}
		\pi(\textbf{u}_c) \propto \exp\left(-\frac{1}{2}\textbf{u}_c^T\mathbf{Q}\textbf{u}_c\right)
	\end{align}
	and corresponding results for $\textbf{u}_r$, $\textbf{u}_s$ and $\textbf{u}_g$.	
	The posterior density of $\textbf{u}_c$ is:
	\begin{equation}\label{five}
		\begin{split}
			\log (\text{posterior})&\propto \log (\text{likelihood})+\log(\text{prior})+\log(\text{hyperprior})\\	
			\log P(\textbf{u}_c|\textbf{y},\boldsymbol\theta,\boldsymbol\gamma) &\propto \log P(\textbf{y}| \textbf{u}_c,\lambda_c,\boldsymbol\theta,\boldsymbol\gamma) + \log P(\textbf{u}_c|\lambda_c,\boldsymbol\theta,\boldsymbol\gamma) + \log P(\log \lambda_c)	\\
			\log P(\textbf{u}_c|\textbf{y},\boldsymbol\theta,\boldsymbol\gamma) &\propto \log P(\textbf{y}| \textbf{u}_c,\boldsymbol\theta,\boldsymbol\gamma) + \log P(\textbf{u}_c|\lambda_c)+ \log P(\log \lambda_c)
		\end{split}
	\end{equation}
	
	Here, $\boldsymbol\theta$ includes the $a_j$, $b_j$, $\beta$ and $e_i$ terms from \eqref{one}. The $\textbf{u}_r$, $\textbf{u}_s$ and $\textbf{u}_g$ terms display the same behaviour as in \eqref{five} with precision parameters $\lambda_r$, $\lambda_s$, $\lambda_g$ in analogous form. These parameters, included in the precision matrix $\textbf{Q}$, will therefore also be different at each level of the hierarchy.
	
	In two dimensions, the behaviour of $\mathbf{Q}$ is more complex than in the one-dimensional case, as its entries are now block-circulant matrices rather than single numbers (\cite{thon2012bayesian}). Moreover, although the order of an IGMRF can be expressed as the rank deficiency of its precision matrix (\cite{rue2005gaussian}), because we now have two variables, we need to impose three constraints, $\sum x_{ij}=0$, $\sum ix_{ij}=0$ and $\sum jx_{ij}=0$, to give a proper joint distribution and a finite marginal distribution. These constraints correspond to the mean of the intercepts, the mean of the first variable's linear trend and the mean of the second variable's linear trend. Each block matrix inside the precision matrix is $ T\times T$ dimensional and therefore the precision matrix is $T^2 \times T^2$ dimensional.  
	\subsection{Precision Matrix on a Torus} \label{Torus1}
\begin{figure}
		\centering
		\includegraphics[scale=0.35]{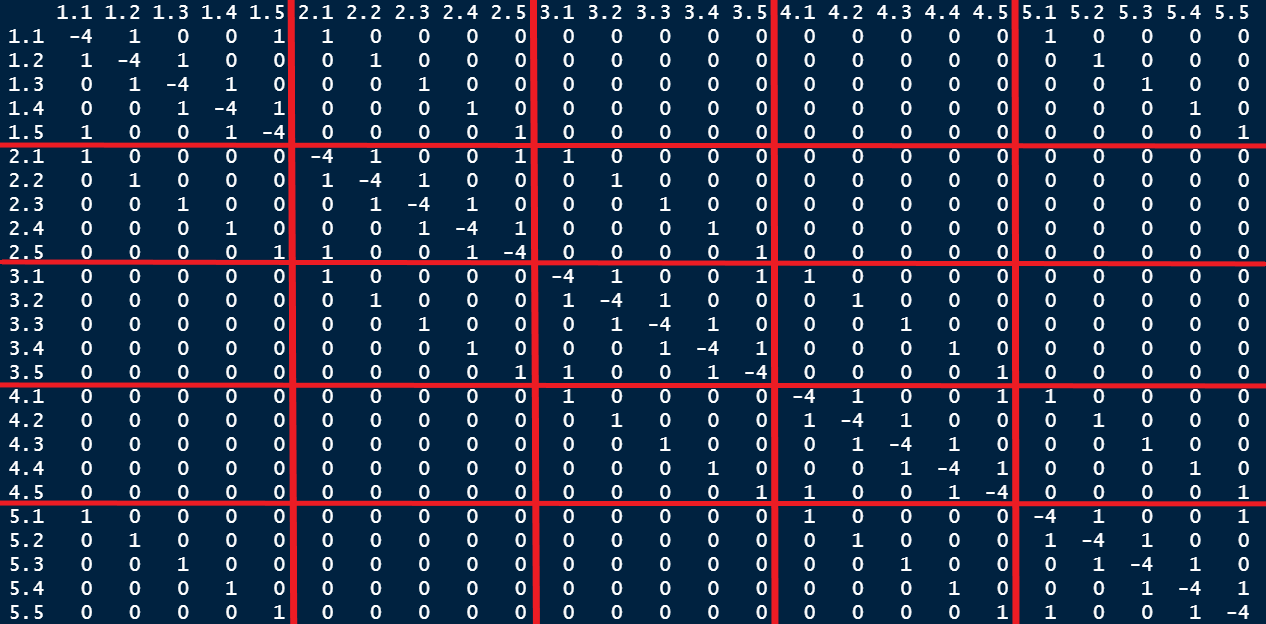}
		\caption{$\mathbf{C}$ matrix for 5 years of SBP and DBP combinations}
		\label{D1}
		\centering
		\includegraphics[scale=0.35]{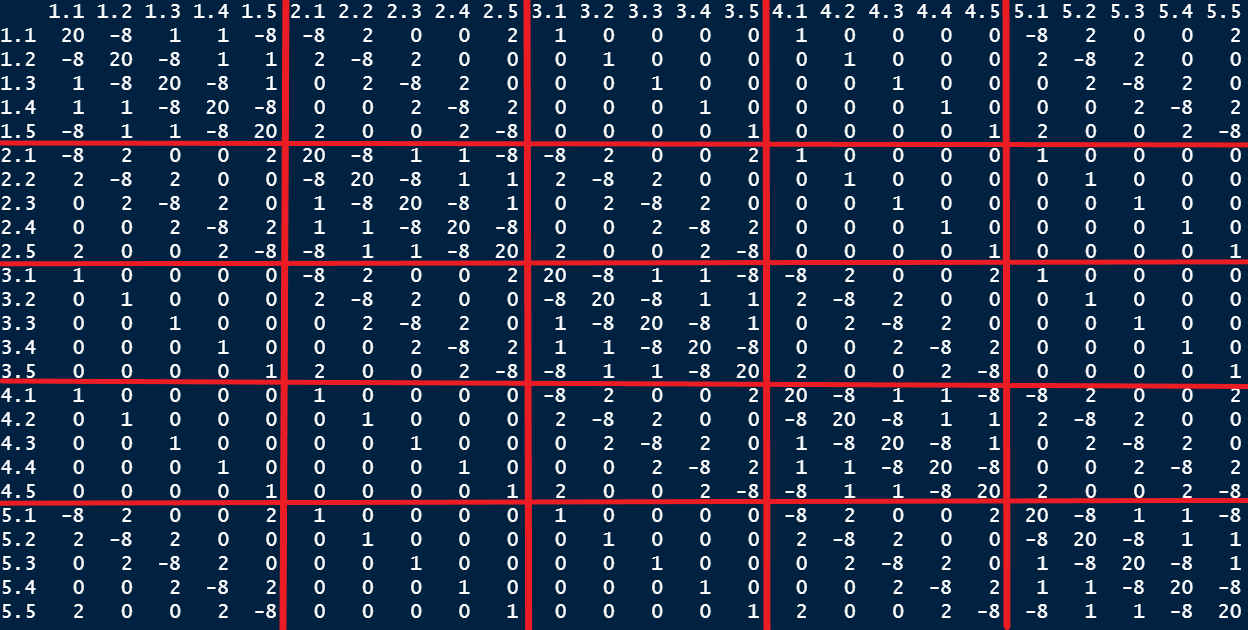}
		\caption{Structure matrix for 5 years of SBP and DBP combinations, $\mathbf{P}=\mathbf{C}^T\mathbf{C}$}
		\label{P1}
	\end{figure}

	The precision matrix for the prior distribution is $\mathbf{Q} = \lambda \times \mathbf{C}^T\mathbf{C}$, where $\mathbf{C}$ is a block-circulant matrix and $\lambda$ is a precision parameter. We consider the case of variables $i$ and $j$ and $T=5$, where at each time point, we wish to make estimates for both variables. We also have a so-called structure matrix, $\mathbf{P} = \mathbf{C}^T\mathbf{C}$. The sub-matrices used for the construction of $\mathbf{C}$ are:
	\begin{align*}
		\mathbf{A}_1=\left( \begin{array}{ccccc}
			-4&1&&&1\\
			1&-4&1&&\\
			&1&-4&1&\\
			&&1&-4&1\\
			1&&&1&-4\\
		\end{array} \right), \hspace{2mm}
		\mathbf{A}_2 = \left( \begin{array}{ccccc}
			1&&&&\\
			&1&&&\\
			&&1&&\\
			&&&1&\\
			&&&&1\\
		\end{array} \right)
	\end{align*}
	with
	\begin{align}\label{matr1}
		\mathbf{C}=\left( \begin{array}{ccccc}
			\mathbf{A_1}&\mathbf{A_2}&\textbf{0}&\textbf{0}&\mathbf{A_2}\\
			\mathbf{A_2}&\mathbf{A_1}&\mathbf{A_2}&\textbf{0}&\textbf{0}\\
			\textbf{0}&\mathbf{A_2}&\mathbf{A_1}&\mathbf{A_2}&\textbf{0}\\
			\textbf{0}&\textbf{0}&\mathbf{A_2}&\mathbf{A_1}&\mathbf{A_2}\\
			\mathbf{A_2}&\textbf{0}&\textbf{0}&\mathbf{A_2}&\mathbf{A_1}\\
		\end{array} \right)
	\end{align}
	Here $\textbf{0}$ represents $T \times T$ zeroes. This matrix shows the behaviour in rows without boundary constraints, with the full matrices shown in Figures \ref{D1} and \ref{P1} respectively.
	
	The construction of the  $\textbf{C}$ matrix is based on forward differences for each variable, with increments:
	\begin{align}\label{delta}
		\Delta^2_{(1,0)}f(D_{i},S_{j}) +\Delta^2_{(0,1)} f(D_{i},S_{j})
	\end{align}
	Here, $\Delta_{(1,0)}$ and $\Delta_{(0,1)}$ are the first-order forward difference operators in the vertical and
	horizontal directions, which can be written as:
	\begin{align}\label{delta2}
		\Delta_{(1,0)}f(D_{i},S_{j})&= f(D_{i},S_{j})-f(D_{i-1},S_{j})\\
		\Delta_{(0,1)}f(D_{i},S_{j})&= f(D_{i},S_{j})-f(D_{i},S_{j-1})
	\end{align}
	Similarly, the second-order forward differences for each variable are defined as:
	\begin{align}\label{delta3}
		\Delta^2_{(1,0)}f(D_{i},S_{j}) =\Delta_{(1,0)}(\Delta_{(1,0)}f(D_{i},S_{j})) = f(D_{i},S_{j})-2f(D_{i-1},S_{j})+f(D_{i-2},S_{j})\\ \Delta^2_{(0,1)}f(D_{i},S_{j})
		=\Delta_{(0,1)}(\Delta_{(0,1)}f(D_{i},S_{j}))=f(D_{i},S_{j})-2f(D_{i},S_{j-1})+f(D_{i},S_{j-2}) 
	\end{align}
	To construct the structure matrix, \textbf{P}, we require the square of the differences in \eqref{delta3} since $\textbf{P}=\textbf{C}^T\textbf{C}$, and so each row of the structure matrix, \textbf{P}, will have increments:
	\begin{equation} \label{2D}
		\begin{split}
			-(\Delta^2_{(1,0)}f(D_{i},S_{j}) &+\Delta^2_{(0,1)} f(D_{i},S_{j}))^2=\\
			&-(\Delta^4_{(1,0)}f(D_{i},S_{j}) +\Delta^4_{(0,1)} f(D_{i},S_{j})+\Delta^2_{(0,1)} \Delta^2_{(1,0)}f(D_{i},S_{j}))
		\end{split}
	\end{equation}
	
	Applying the forward differences in \eqref{2D}, we have the following entries for each row, based on the nearest neighbours for each $(i,j)$ combination:
	\begin{equation}\label{s}
		\begin{split}
			f(D_{i-2},S_{j})&+2f(D_{i-1},S_{j-1})-8f(D_{i-1},S_{j})+2f(D_{i-1},S_{j+1})+f(D_{i},S_{j-2})\\
			&-8f(D_{i},S_{j-1})+20f(D_{i},S_{j})-8f(D_{i},S_{j+1})+f(D_{i},S_{j+2})\\&+2f(D_{i+1},S_{j-1})-8f(D_{i+1},S_{j})+2f(D_{i+1},S_{j+1})+f(D_{i+2},S_{j})
		\end{split}	
	\end{equation}
	This result leads to the structure matrix presented in Figure \ref{P1}, which has a torus form (\cite{thon2012bayesian}), where nodes on opposing edges are treated as neighbours. This is not appropriate for our analyses given that time is linear, and so we require an alternative specification with boundary constraints. An isotropic precision matrix that includes the corners $x_{1,1}, x_{1,n}, x_{n,1}, x_{n,n}$ omitted above is available (\cite{rue2005gaussian}), but this is also on a torus (\cite{rue2005gaussian}, \cite{thon2012bayesian}) and so is not appropriate.
	\subsection{Precision Matrix with Boundary Constraints}\label{Bounds2}
	A structure matrix $\textbf{P}$ that is not on a torus is available (\cite{yue2010nonstationary}), and this more realistic specification has been used in our analyses of blood pressure. Here, we define the block matrices of $\textbf{P}$ such that each row and column sums to zero, taking into account only data in neighbouring years. We use sub-matrices:
	\begin{align*}
		\mathbf{A}_1=\left( \begin{array}{ccccc}
			6&-5&1&&\\
			-5&12&-6&1&\\ 
			1&-6&12&-6&1\\
			&1&-6&12&-5\\
			&&1&-5&6\\
		\end{array} \right),\hspace{2mm} 
		\mathbf{A}_2=\left( \begin{array}{ccccc}
			-5&2&&&\\
			2&-7&2&&\\
			&2&-7&2&\\
			&&2&-7&2\\
			&&&2&-5\\
		\end{array} \right)
	\end{align*}
	$$\mathbf{A_3}=\textrm{diag}(1,1,1,1,1)$$
	\begin{align*}
		\mathbf{A}_4=\left( \begin{array}{ccccc}
			12&-7&1&&\\-7&20&-8&1&\\1&-8&20&-8&1\\
			&1&-8&20&-7\\&&1&-7&12\\
		\end{array} \right), \hspace{2mm}
		\mathbf{A}_5=\left( \begin{array}{ccccc}
			-6&2&&&\\2&-8&2&&\\&2&-8&2&\\
			&&2&-8&2\\&&&2&-6\\
		\end{array} \right)
	\end{align*}
	This gives a precision matrix: 
	\begin{align}\label{yue}
		\lambda \mathbf{P}=\lambda\left( \begin{array}{ccccc}
			\textbf{A}_1&\textbf{A}_2&\textbf{A}_3&&\\
			\textbf{A}_2&\textbf{A}_4&\textbf{A}_5&\textbf{A}_3&\\ 
			\textbf{A}_3&\textbf{A}_5&\textbf{A}_4&\textbf{A}_5&\textbf{A}_3\\
			&\textbf{A}_3&\textbf{A}_5&\textbf{A}_4&\textbf{A}_2\\
			&&\textbf{A}_3&\textbf{A}_2&\textbf{A}_1\\
		\end{array} \right)
	\end{align}
	This matrix has both overall boundary constraints and constraints within the block matrices, with a rank of $T^2-3$ based on the $\sum x_{ij}=0$, $\sum ix_{ij}=0$ and $\sum jx_{ij}=0$ constraints described above. The matrix is shown in detail in Figure 3, where for the case $T=5$ only the $13^{th} $ row has no boundary constraints. 
	\begin{figure}
		\centering
		\includegraphics[scale=0.34]{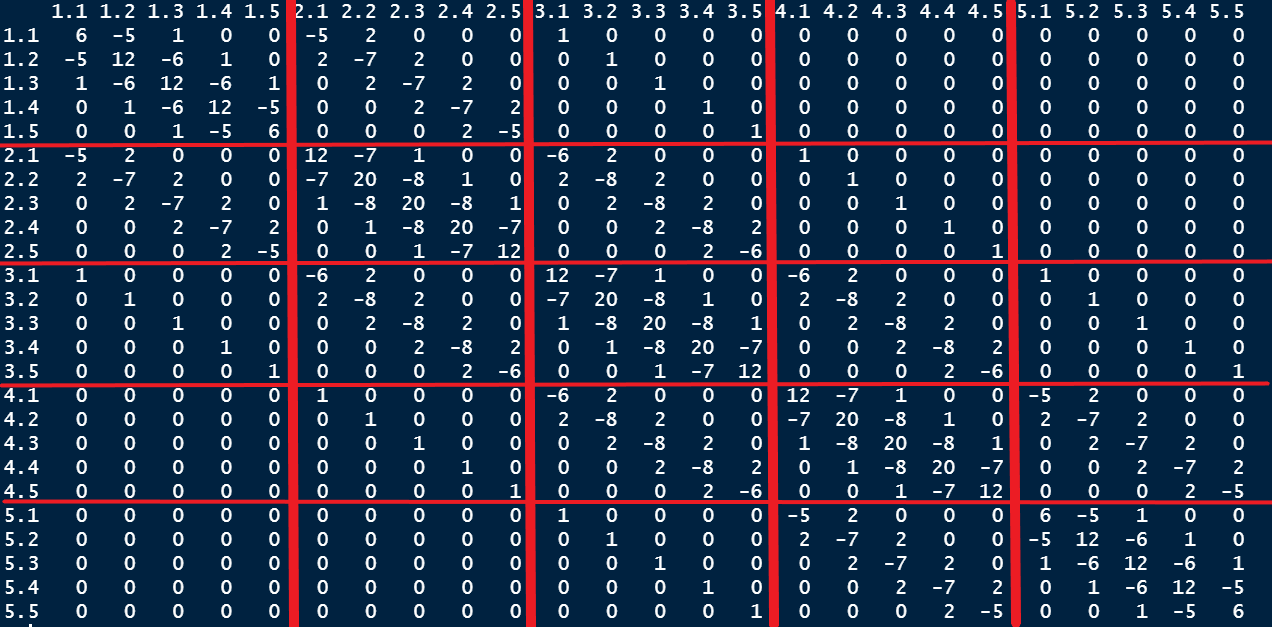}
		\caption{Structure matrix for 5 years of SBP and DBP combinations, $\textbf{P}= \textbf{C}^T\textbf{C}$}
		\label{fig:P4}
	\end{figure}    
	\begin{figure}
		\centering
		\includegraphics[scale=0.3]{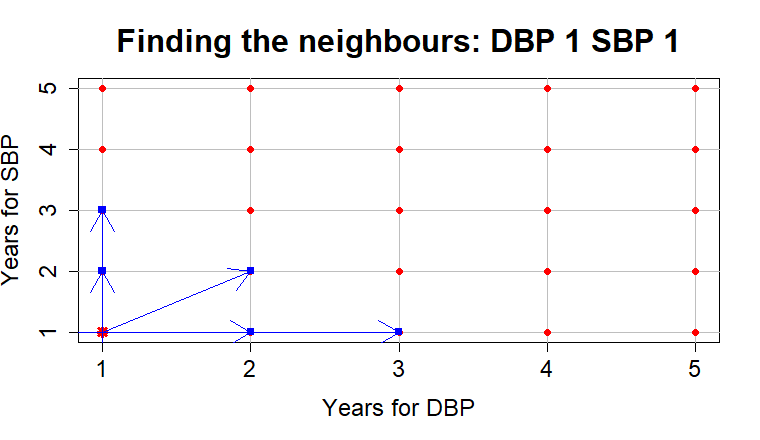}
		\includegraphics[scale=0.3]{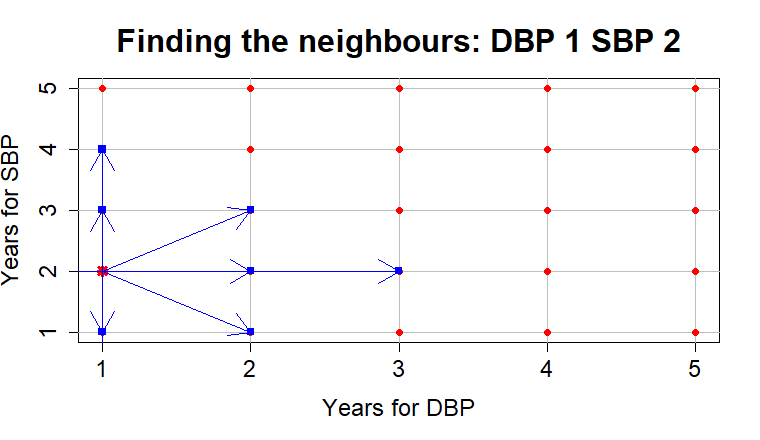}\\
		\includegraphics[scale=0.3]{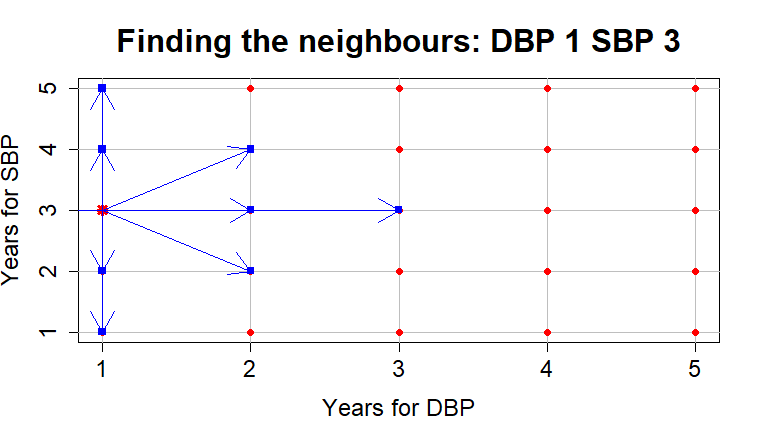}
		\includegraphics[scale=0.3]{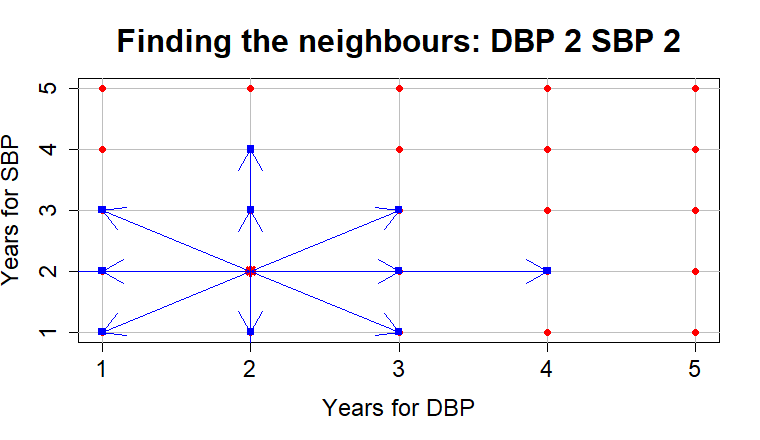}\\
		\includegraphics[scale=0.3]{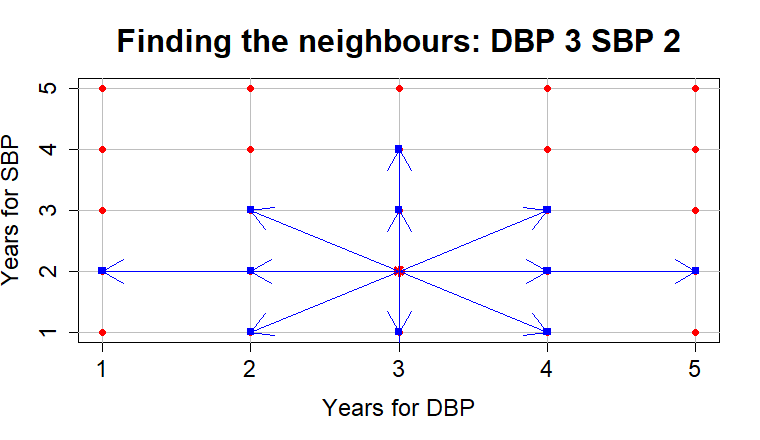}
		\includegraphics[scale=0.3]{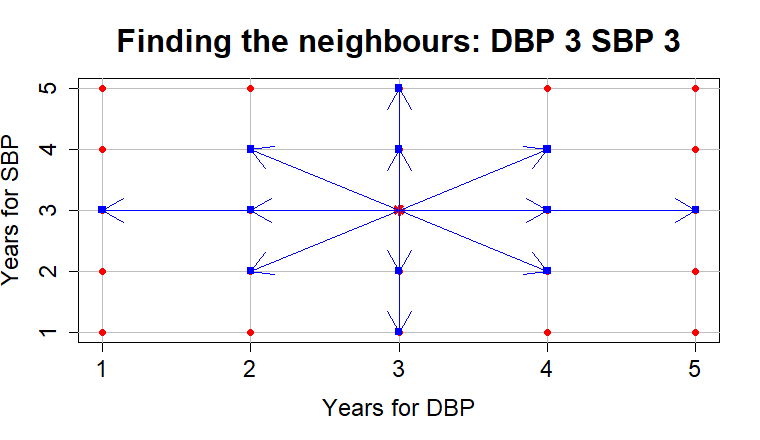}
		\caption{Neighbouring years for DBP and SBP in the case $T=5$}
		\label{fig:fig4}
	\end{figure}
	
	Each cell of the precision matrix $\lambda\mathbf{P}$ represents the inverse of the covariance between DBP and SBP in different years, except for the diagonals, which provide the inverse of the variance for DBP and SBP in the same year. These relationships are shown in detail in Figure 4, where the number of neighbours depends on the combination of years for which DBP and SBP are being estimated. For example, when both DBP and SBP variables are at $(i,j)=(1,1)$, only information from the years ahead is available, with the opposite effect when $(i,j)=(5,5)$. For a two-dimensional second-order random walk, we consider neighbours up to two steps in each direction (i.e., forward or backward either vertically or horizontally), where moving one step diagonally comprises two steps, one vertical and one horizontal. The numbers of possible neighbours are 6, 8, 9, 11 or 12 at the boundaries and 13 elsewhere. The differences are calculated taking into account this behaviour at the boundaries (\cite{yue2010nonstationary}):
	\begin{equation}\label{distYue}
		\begin{split}
			&\sum_{i=2}^{T-1}\sum_{j=2}^{T-1} \lbrace\Delta_0^{2}f(D_i,S_j)\rbrace^2+ \lbrace\Delta_1f(D_1,S_1)\rbrace^2+
			\lbrace\Delta_2f(D_T,S_1)\rbrace^2\\
			&+\lbrace\Delta_3f(D_1,S_T)\rbrace^2+ \lbrace\Delta_4f(D_T,S_T)\rbrace^2+
			\sum_{i=2}^{T}(\lbrace\Delta_5f(D_i,S_1)\rbrace^2+
			\lbrace\Delta_6f(D_i,S_T)\rbrace^2)\\
			&+\sum_{j=2}^{T}(\lbrace\Delta_7f(D_1,S_j)\rbrace^2+
			\lbrace\Delta_8f(D_T,S_j)\rbrace^2)
		\end{split}	
	\end{equation}
	
	Here, each of the forward differences can be written as:
	\begin{align*}
		\Delta_0^{2}f(D_i,S_j)&=\\
		(\Delta^{2}_{(1,0)}+\Delta^{2}_{(0,1)})f(D_{i},S_{j})&=\Delta^{2}_{(1,0)}f(D_{i},S_{j})+\Delta^{2}_{(0,1)}f(D_{i},S_{j})\\
		\lbrace\Delta_0^{2}f(D_i,S_j)\rbrace^2&=\\
		(\Delta^{2}_{(1,0)}+\Delta^{2}_{(0,1)})^{2}f(D_{i},S_{j})
		&=(\Delta^{4}_{(1,0)}+\Delta^{4}_{(0,1)}+2\Delta^{2}_{(1,0)}\Delta^{2}_{(0,1)})f(D_{i},S_{j})
	\end{align*}
	\begin{align*}
		\lbrace\Delta_1f(D_1,S_1)\rbrace^2&=(\Delta_{(1,0)}+\Delta_{(0,1)})^2f(D_1,S_1)\\
		\lbrace\Delta_2f(D_T,S_1)\rbrace^2&= (\Delta_{(1,0)}+\Delta_{(0,1)})^2f(D_T,S_1)\\
		\lbrace\Delta_3f(D_1,S_T)\rbrace^2
		&= (\Delta_{(1,0)}
		+\Delta_{(0,1)})^2f(D_1,S_T)\\
		\lbrace\Delta_4f(D_T,S_T)\rbrace^2 &=(\Delta_{(1,0)}+\Delta_{(0,1)})^2f(D_T,S_T)\\
		\lbrace\Delta_5f(D_i,S_1)\rbrace^2&= (\Delta_{(1,0)}^{2}+\Delta_{(0,1)})^2f(D_{i},S_1)\\
		\lbrace\Delta_6f(D_i,S_T)\rbrace^2&= (\Delta_{(1,0)}^{2}+\Delta_{(0,1)})^2f(D_{i},S_T)\\
		\lbrace\Delta_7f(D_1,S_j)\rbrace^2&= (\Delta_{(1,0)}+\Delta_{(0,1)}^{2})^2f(D_1,S_{j})\\
		\lbrace\Delta_8f(D_T,S_j)\rbrace^2&=
		(\Delta_{(1,0)}+\Delta_{(0,1)}^{2})^2f(D_T,S_{j})
	\end{align*}
	When $T=5$, for each variable we have $(5-2)^2 + 4 + (5-2)*2 + (5-2)*2 = 25$ combinations of differences, each of which has different behaviour depending on its position. For example, in \eqref{yue} the middle row of the precision matrix is:
	\begin{equation}\label{centr}	
		\begin{split}
			f(D_{i-2},S_{j})&+2f(D_{i-1},S_{j-1})-8f(D_{i-1},S_{j})+2f(D_{i-1},S_{j+1})+f(D_{i},S_{j-2})\\
			&-8f(D_{i},S_{j-1})+20f(D_{i},S_{j})-8f(D_{i},S_{j+1})+f(D_{i},S_{j+2})\\&+2f(D_{i+1},S_{j-1})-8f(D_{i+1},S_{j})+2f(D_{i+1},S_{j+1})+f(D_{i+2},S_{j})
		\end{split}		
	\end{equation}
	
	The precision matrix in \eqref{yue} is an IGMRF of order $k=2$ in dimensions $d=2$ which is an improper GMRF of rank $T^2-3$, where there are three constraints and so three eigenvalues of $\mathbf{Q}$ with value zero (\cite{paciorek2009technical}).
	
	Finally, the prior is a stationary GMRF. This means that if $x_{-ij}$ denotes the full field $x$ with sites $i$ and $j$ excluded, the full conditionals $\pi(x_{ij}|x_{-ij})$ have constant parameters not depending on $ij$, such that the local properties are the same across all years.
	
	\section{Implementation using Metropolis within Gibbs}
	\subsection{Block-sampling Metropolis-Hastings algorithm}
	For complex Bayesian hierarchical models such as this, block updating can be used to avoid slow convergence and poor mixing caused by strong dependence between parameters and their hyper-parameters (\cite{rue2005gaussian}), using a joint proposal and hence a joint posterior distribution. In our case, we use block updating for the parameters and hyper-parameters of the linear and non-linear models. In this section, we describe block updating for the non-linear model. 
	
	\subsection{Full conditional distributions}
	Using a two-dimensional non-linear model, the interaction is already integrated implicitly in the model by $\textbf{P}$ as shown in \eqref{yue},  for each $\textbf{u}_c$, $\textbf{u}_r$, $\textbf{u}_s$, $\textbf{u}_g$. The non-linear term for country $j$ and study $i$, $\mathbf{u_{j[i]}}$, present in the likelihood in \eqref{one} and \eqref{norm}, can be represented as: 
	\begin{align}\label{nonlin1}
		\boldsymbol{u}_{j} &=\boldsymbol{u}_{j}^c+\boldsymbol{u}_{k[j]}^r+\boldsymbol{u}_{l[j]}^s+\boldsymbol{u}^g
	\end{align}
	In matrix form, the non-linear model for all country-year combinations can be written as:
	\begin{align}
		\boldsymbol{Mu}= \textbf{M}^{c}\boldsymbol{u}^{c} + \textbf{M}^{r}\boldsymbol{u}^{r} +
		\textbf{M}^{s}\boldsymbol{u}^{s} + \textbf{M}^{g}\boldsymbol{u}^{g}\label{nonlin2}
	\end{align}
	
	Here, $\boldsymbol{M}^c$ is a translation matrix of $I\times J*T^2$ dimensionality that assigns the elements of $\textbf{u}^c$ to studies observed in particular country-year combinations. The $\textbf{M}^r$, $\textbf{M}^s$ and $\textbf{M}^g$ matrices correspond to region-year,  super-region-year and globe-year combinations, and we have prior distributions
	\begin{align}\label{prior1}
		\textbf{u}_{j}^c \sim \mathcal{N}(\textbf{0},\lambda_c\textbf{P}),
		\hspace{2mm}\textbf{u}_{k[j]}^r \sim \mathcal{N}(\textbf{0},\lambda_r\textbf{P}),
		\hspace{2mm}\textbf{u}_{l[j]}^s \sim \mathcal{N}(\textbf{0},\lambda_s\textbf{P}),
		\hspace{2mm}\textbf{u}^g \sim \mathcal{N}(\textbf{0},\lambda_g\textbf{P})
	\end{align}
	
	Here, \textbf{0} is a vector of length $T^2$, $\textbf{P}$ is a $T^2\times T^2$ matrix, and each of the non-linear terms is a vector that comprises all possible combinations of years for DBP and SBP: for instance, for the country-level non-linear term, we have a vector of length $T^2$, $\textbf{u}_{j}^c=(u_{11},u_{12},\dots, u_{32},u_{34},u_{35},\dots, u_{TT})$ for a specific country. The precision parameters, $\lambda_c, \lambda_r, \lambda_s, \lambda_g$, are each scalars.
	
	While the model allows estimation for $T^2$ combinations, we can only add information to the likelihood when data are present, and this is only the case when both DBP and SBP are in the same year because the measurements are taken for individuals simultaneously. Here, we describe the posterior at the country level of the hierarchy, but the posteriors for the other levels are analogous. For $\textbf{u}^c$, we have:
	\begin{equation} 
		\begin{split}   
			\log(P(\textbf{u}^c,\lambda_c|\textbf{y},\boldsymbol{\theta},\gamma)) &\propto (\textbf{y}-\textbf{F}(\textbf{Z}\boldsymbol{\phi}+1)\boldsymbol{\theta}-\textbf{M}_c(\textbf{Z}\boldsymbol{\phi}+1)\textbf{u}^c-\textbf{Z}\boldsymbol{\psi}-(\textbf{C}_c\textbf{Z})1_5)^{T}\boldsymbol{\Sigma^{-1}}\\ 
			&(\textbf{y}-\textbf{F}(\textbf{Z}\boldsymbol{\phi}+1)\boldsymbol{\theta}-\textbf{M}_c(\textbf{Z}\boldsymbol{\phi}+1)\textbf{u}^c-\textbf{Z}\boldsymbol{\psi}-(\textbf{C}_c\textbf{Z})1_5) \\+ &\log(P(\textbf{u}^c|\lambda_c))+ \log P(\log \lambda_c)
		\end{split}
	\end{equation} 
	For the posterior of the non-linear model, we will keep only the $\textbf{u}^c$ and $\textbf{M}_c$ terms that comprise the non-linear model, and we can state the posterior distribution using canonical parameterisation, $f(\textbf{u}^c,\lambda_{u^c}|\boldsymbol\mu_{u^c},\textbf{Q}_{u^c})\sim N(\textbf{Q}_{u^c}\boldsymbol\mu_{u^c}, \textbf{Q}_{u^c})$. Therefore, the distribution of $\textbf{u}^c$ is:
	\begin{equation}\label{canon}
		\begin{split}
			\textbf{Q}_{u^c} &= (\textbf{M}_c*(\textbf{Z}\phi+1))^T\boldsymbol{\Sigma^{-1}}(\textbf{M}_c*(\textbf{Z}\phi+1))+ I_j\otimes\lambda_{c}\textbf{P}\\
			\textbf{Q}_{u^c}\boldsymbol\mu_{u^c} &= (\textbf{M}_c*(\textbf{Z}\phi+1))^{T}\boldsymbol{\Sigma^{-1}}[\textbf{y}-\textbf{F}(\textbf{Z}\phi+1)\boldsymbol{\theta}-(\textbf{Z}\psi-((\textbf{C}c)\textbf{Z})1_5)]
		\end{split}
	\end{equation}
	where $\textbf{Z}$ expresses the age data, $\boldsymbol{\psi}$ expresses the constant term, $\boldsymbol{\phi}$ is the mean term and $\textbf{C}$ are country-specific random effects in the age model.
	\subsection{Defining constraints}\label{constr}
	To construct the posterior and define all the parameters, we require data that correspond to the linear constraints. Again, while we refer only to the country-level case, the same applies for the other levels. For the hyper-prior, we propose $\log\lambda_c^*$  from a $\mathcal{N}(\log\lambda_c^{i-1},\omega^2_{\log\lambda_c})$ distribution, while $\textbf{Q}_{u^c}$ in \eqref{canon} is a block diagonal which refers to full-conditional within-country correlation. Each block corresponds to a country and the correlations that exist between the years of data for a country. Therefore, we have $J$ blocks corresponding to $J$ countries each containing $T^2$ year combinations. In addition, we will have $J$ possible values of $\boldsymbol{u}_j^c$, where $\boldsymbol{u}_j^c$ is a vector of $T^2$ dimensions. For the block corresponding to a country $j$, there are four possible scenarios: 
	\begin{enumerate}
		\item Country $j$ has no data\\
		The likelihood does not contribute, hence, $\textbf{Q}_{u^c}$ in \eqref{canon} depends solely on the prior, so we propose $\textbf{u}_j^{c*} \sim \mathcal{N}(0,(\lambda_c \textbf{P})^{-1})$. The rank of \textbf{P} is $T^2-3$, corresponding to infinite prior variance on the mean (intercept) and the two linear trends of $\textbf{u}_j^c$. We constrain these three linear combinations of $\textbf{u}^c_{j}$ to zero by taking the generalised inverse of $\textbf{P}$, setting  the last three eigenvalues of $\textbf{P}$ to $\infty$.
		
		\item Country $j$ has one year of data\\
		$\textbf{Q}_{u^c}$  has rank  $T^2-2$ because the mean (intercept) is defined by the data but neither of the linear trends of $\textbf{u}_j^c$ are. Hence, we constrain the linear trends of $\textbf{u}_j^c$ to zero by taking the generalised inverse of $\textbf{Q}_{u^c}$, setting $\textbf{Q}_{u^c}$'s last two eigenvalues to $\infty$. For identifiability reasons we constrain the mean of $\textbf{u}_j^c$'s intercept to zero. 
		
		\item Country $j$ has two years of data\\
		$\textbf{Q}_{u^c}$  has rank  $T^2-1$ because the mean and  the first linear trend of $\textbf{u}_j^c$ are defined by the data. Hence, we constrain the second linear trend $\textbf{u}_j^c$ to zero by taking the generalised inverse of $\textbf{Q}_{u^c}$, setting $\textbf{Q}_{u^c}$'s last eigenvalue to $\infty$. For identifiability reasons we constrain the mean of $\textbf{u}_j^c$'s intercept and the mean of the first $\boldsymbol{u}_j^c$'s linear trend to zero.

		\item Country $j$ has three or more years' data\\
		$\textbf{Q}_{u^c}$  has rank  $T^2$ because the mean and both of the linear trends of $\textbf{u}_j^c$ are identified by the data. Again, for identifiability reasons we  constrain the mean of $\boldsymbol{u}_j^c$'s intercept and both $\boldsymbol{u}_j^c$'s linear trends to zero.
		
	\end{enumerate}
	We accept $\textbf{u}^{c*}$ and $\lambda_c^*$ with probability $r$:
	\begin{align} \label{ratio}
		r=\dfrac{P(\textbf{y}|\textbf{u}^{c*},.)P(\textbf{u}^{c*}|\lambda^*_c)P(\log\lambda^*_c)P(\textbf{u}^{c(i-1)}|\textbf{A}\textbf{u}_j^{c(i-1)}=0               \,\forall j, \lambda_c^{i-1},\textbf{y},.) }{P(\textbf{y}|\textbf{u}^{c(i-1)},.)P(\textbf{u}^{c(i-1)}|\lambda^{(i-1)}_c)P(\log\lambda^{(i-1)}_c)P(\textbf{u}^{c*}|\textbf{A}\textbf{u}_j^{c*}=0 \, \forall j, \lambda_c^{i-1},\textbf{y},.)} 
	\end{align} 
	Here, $\textbf{A}$ is a $3\times T^2$ constraint matrix, where the first row is a vector $\mathbf{1}$, the second row is a vector of centred time values for the DBP variable, and the third row is a vector of centred time values for the SBP variable.
	\section{Simulation studies}
	To test the model and its implementation, we simulated data from 61 countries using the NCD Risk Factor Collaboration regional classification for blood pressure (\cite{zhou2017worldwide}): three countries each from 20 regions, and Japan in high-income Asia Pacific. We simulated SBP and DBP data for five years for each country (i.e., the fourth scenario in section \ref{constr}). Nonetheless, there was still substantial sparsity, since we did not have data for all the year combinations. We split the data into training and test sets, such that most countries would be represented in the training set. As Japan was the only country in its region, it was included in each training set. 
	
	We carried out five folds of cross-validation, with the average coverage and errors shown for super-regions in Table 1. The coverage of actual uncertainty by estimated credible intervals was generally more than 95\% (i.e. conservative), although with cases where the coverage was somewhat less (i.e., anti-conservative). As shown in Table 1, the errors were small for both DBP and SBP, while the errors are difficult to interpret for the interaction terms.
	
	\begin{table}
		\caption{Coverage and error in cross-validation}
		\begin{centering}
			\begin{tabular}{lrrrrrr}
				\multicolumn{1}{c}{} &
				\multicolumn{3}{c}{Coverage} &
				\multicolumn{3}{c}{Error}\\ 
				Regions&DBP&SBP&Int&DBP&SBP&Int\\
				\hline
				Total&0.984&0.986&0.948&0.02&-0.01&94.8\\
				Central and Eastern Europe&0.994&1&0.952&-0.06&-0.004&298.3\\
				C. Asia, Middle East \& N. Africa&0.990&0.987&0.865&-0.02&-0.18&1119.4\\
				East and South East Asia&0.955&0.997&0.897&0.08&0.02&341.5\\
				High-income Western countries&0.998&0.994&0.996&-0.01&-0.1&902.8\\
				Latin America and Caribbean&0.978&0.982&1&0.09&0.02&-1041.9\\
				Oceania&0.970&1&0.955&0.11&-0.03&57.2\\
				South Asia&0.990&0.983&0.830&-0.10&-0.11&1245.1\\
				Sub-Saharan Africa&0.990&0.982&0.954&0.02&0.77&-444.0\\
			\end{tabular}\hfill
		\end{centering}
		\label{tab:tab1}
	\end{table}
	
	\begin{figure}
		\includegraphics[scale=0.4]{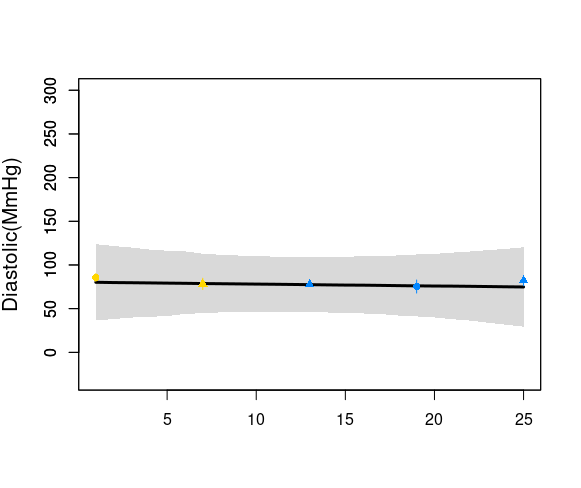}
		\includegraphics[scale=0.4]{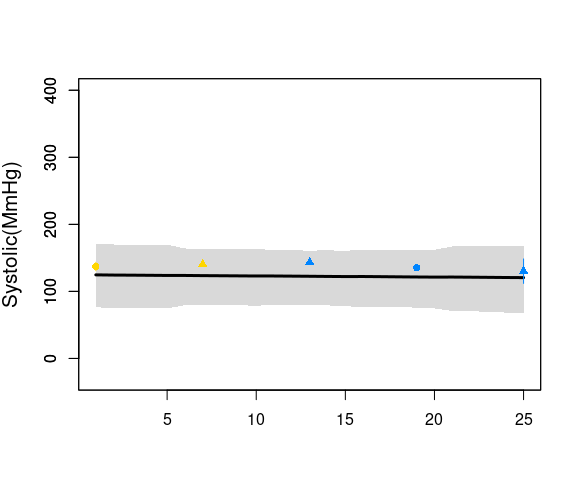}\\
		\includegraphics[scale=0.4]{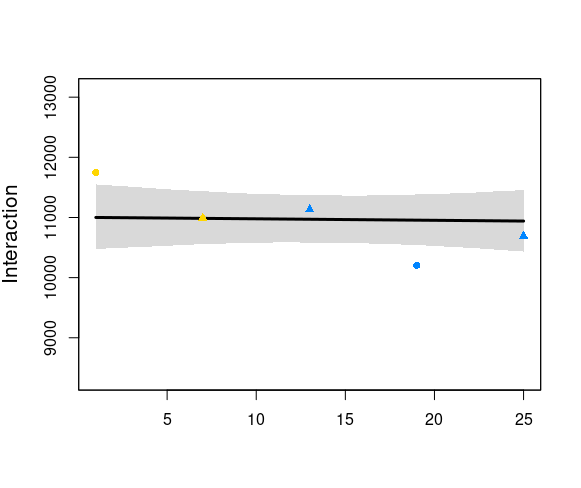}
		\includegraphics[scale=0.5]{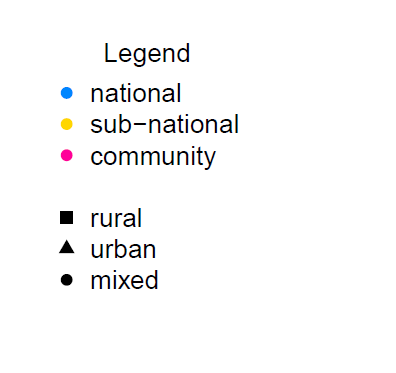}
		\caption{\label{fig:vanuatu}Fit plots for 50-year old females in Vanuatu from cross validation. The solid line represents the posterior mean and the shaded area the 95\% credible interval.}
	\end{figure}
	
	Figure \ref{fig:vanuatu} shows an example of model fitting, for 50-year-old females in Vanuatu, demonstrating that the algorithm functions well. The x-axis has values from 1 to 25 that encode the 25 possible combinations for five years of DBP data and five years of SBP data. Data are only simulated for five of these combinations, as described above, but this allows us to make estimates for the other possible combinations.

	\section{Application to real data}
	We downloaded publicly available data from the WHO website for 38 countries in the period 2004 to 2014. The precision matrix described in \eqref{yue} was therefore $11^2 \times 11^2$ dimensional in this case. While this increased the time required for the MCMC algorithm to converge and mix, it was still tractable using a high-performance computing system. We used the regional classification described above, making estimates for 48 countries in total, including cases where countries had data for zero, one, two or more years, i.e., each of the cases described previously. Figure 6 shows an example of the model fit for Myanmar, where the interaction between SBP and DBP is greatest where these values are both lowest, while we observed an opposite effect in Indonesia, for example. Neither of these effects would have been captured by the models used in previous estimation of blood pressure trends.
	\begin{figure}[h]
		\includegraphics[scale=0.7]{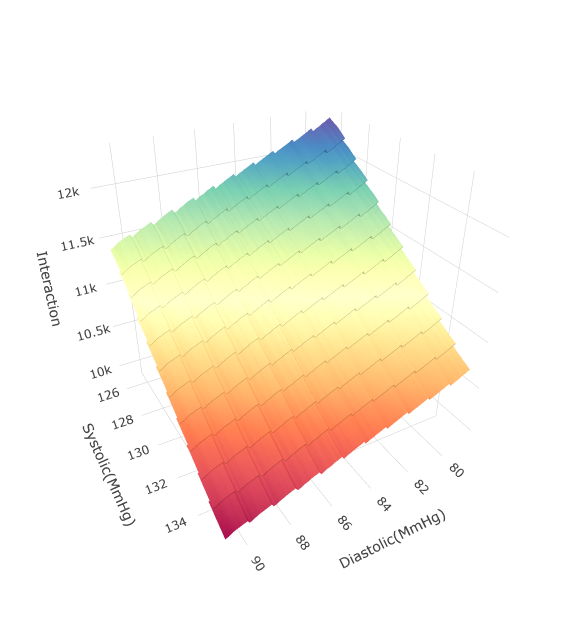}
		\caption{\label{fig:Myanmar1}Estimates for 50-year old males in Myanmar from 2004 to 2014.}
	\end{figure}
	
	\section{Summary and future work}
	We have shown that it is possible to fit a model to blood pressure data in two dimensions, and in particular that is possible to fit a non-linear surface that implicitly estimates the interaction between these variables. Our practical implementation will be of interest to any researchers with sparse bivariate data.
	
	Given that the hyper-parameters needs to be parameterised each time the number of nodes or the order of the IGMRF that is taken into account is changed, we plan to explore the use of penalised complexity (PC) priors. We are also interested in whether this model can be fitted using alternatives to MCMC.\\ 
	
	\noindent\textbf{Acknowledgements} Specialist and High Performance Computing systems were provided by Information Services at the University of Kent.
	
	\bibliographystyle{apacite}
	\bibliography{bibl}	
\end{document}